\documentclass[prl,final,twocolumn,superscriptaddress,aps]{revtex4-2}
\usepackage{amsmath}
\usepackage{amssymb}
\usepackage[utf8]{inputenc}
\usepackage[colorlinks=True,linkcolor=blue,citecolor=blue,urlcolor=blue]{hyperref}
\usepackage{xcolor}
\usepackage{graphicx}

\usepackage{dsfont}

\date{\today}

\definecolor{Ured}{HTML}{cc0000}
\definecolor{Ublue}{HTML}{1f65cf}
\definecolor{Ugreen}{HTML}{198a11}

\newcommand{\eye}{\mathds{1}}
\newcommand{\mat}[1]{\mathsf{#1}}
\DeclareMathOperator{\Tr}{Tr}

\begin{document}
\title{Random matrix theory for quantum and classical metastability in local Liouvillians}
\author{Jimin L. Li}
\affiliation{Department of Chemistry, University of Cambridge, Lensfield Road, Cambridge, CB2 1EW, UK}
\author{Dominic C. Rose}
\affiliation{School of Physics and Astronomy, University of Nottingham, Nottingham, NG7 2RD, UK}
\affiliation{Centre for the Mathematics and Theoretical Physics of Quantum Non-Equilibrium Systems, University of Nottingham, Nottingham, NG7 2RD, UK}
\affiliation{Department of Physics and Astronomy, University College London, Gower Street, London WC1E 6BT, UK}
\author{Juan P. Garrahan}
\affiliation{School of Physics and Astronomy, University of Nottingham, Nottingham, NG7 2RD, UK}
\affiliation{Centre for the Mathematics and Theoretical Physics of Quantum Non-Equilibrium Systems, University of Nottingham, Nottingham, NG7 2RD, UK}
\author{David J. Luitz}
\affiliation{Max Planck Institute for the Physics of Complex Systems, Noethnitzer Str. 38, 01187 Dresden, Germany}

\begin{abstract}
We consider the effects of strong dissipation in quantum systems with a notion of locality, which induces a hierarchy of many-body relaxation timescales as shown in [\href{https://journals.aps.org/prl/abstract/10.1103/PhysRevLett.124.100604}{Phys. Rev. Lett. \textbf{124}, 100604 (2020)}]. If the strength of the dissipation varies strongly in the system, additional separations of timescales can emerge, inducing a manifold of metastable states, to which observables relax first, before relaxing to the steady state. Our simple model, involving one or two ``good'' qubits with dissipation reduced by a factor $\alpha<1$ compared to the other ``bad'' qubits, confirms this picture and admits a perturbative treatment.
\end{abstract}

\maketitle

\textbf{Introduction --- }Quantum many-body systems are generically complex, and obtaining an analytic understanding of the position of all spectral resonances is often hopeless. It was realized early on \cite{wigner_characteristic_1955,wigner_random_1967,dyson_statistical_1962,dyson_statistical_1962-1,dyson_statistical_1962-2,mehta_random_2004} that this complexity is in fact so great that many statistical properties of the spectrum are identical with those of random matrices sampled from an ensemble determined by the symmetry of the system. These pioneering observations have been subsequently refined, resulting in cornerstones of our understanding of thermalization in unitary quantum many-body systems by virtue of the eigenstate thermalization hypothesis \cite{feingold_ergodicity_1984,deutsch_quantum_1991,deutsch_eigenstate_2018,srednicki_chaos_1994,srednicki_thermal_1996,rigol_thermalization_2008,dalessio_quantum_2016,borgonovi_quantum_2016}, only with exceptions in integrable \cite{berry_level_1977,kinoshita_quantum_2006,rigol_relaxation_2007,langen_experimental_2015}, many-body localized \cite{anderson_absence_1958, basko_metalinsulator_2006,oganesyan_localization_2007,znidaric_many-body_2008,pal_many-body_2010,luitz_many-body_2015,nandkishore_many-body_2015,abanin_colloquium_2019,luitz_anomalous_2016,luitz_ergodic_2017,abanin_recent_2017}, time-crystalline \cite{khemani_phase_2016,else_floquet_2016,khemani_brief_2019} or scarred and constrained systems \cite{bernien_probing_2017,turner_weak_2018,Pancotti2020}.

This thinking was recently pushed to the realm of open quantum systems, with random matrix models of Markovian dissipation defined via random Liouvillians \cite{denisov_universal_2019,sa_spectral_2019,can_random_2019,can_spectral_2019}, revealing fascinating spectral features of generic purely dissipative systems, in particular a spectral support which has the shape of a ``lemon'' \cite{denisov_universal_2019,sa_spectral_2019}, much different from the circular spectrum of non-Hermitian Ginibre random matrices \cite{ginibre_statistical_1965}. This feature is also present in classical master equations, where typical transition rate matrices have a similar spectral support \cite{timm_random_2009,tarnowski_random_2021}.

Such random matrix models of open quantum many-body systems represent the behavior of typical systems, rather than of a specific model. While they reproduce global properties of more realistic, microscopic models, they miss a crucial ingredient: the locality of (dissipative) interactions. It was recently shown that random matrix models for local Liouvillians can be devised exhibiting a hierarchy of relaxation timescales \cite{wang_hierarchy_2020}. These models limit the jump operators in the Lindblad equation to low complexity Pauli strings, thus encoding few-body interactions. In the absence of detailed microscopic knowledge, this accurately models dissipation in current quantum computer prototypes, and the predicted timescales were in fact observed experimentally on the IBM platform \cite{sommer_many-body_2021}.

Here we apply such a local random matrix model approach to  systems with strongly varying dissipation. We are specifically interested in the appearance of metastable states due to a separation of timescales caused by fast and slow dissipation modes in the system, which we model by the existence of good qubits with low dissipation rates in a system of otherwise bad qubits where dissipation is fast.
In this setup a \emph{metastable manifold} (MM) emerges \cite{macieszczak_towards_2016}, to which the dynamics starting from an arbitrary initial state relaxes quickly. At intermediate times, the dynamics is effectively restricted to the MM, before eventual relaxation to the global steady state at long times. We argue that this model contains the essence of the physics to be expected in a quantum computer with good and bad qubits and is furthermore the simplest generic model to study metastability. Our model generalizes findings of MMs in the presence of local loss terms \cite{froml_fluctuation-induced_2019,wolff_nonequilibrium_2020}.

\textbf{Model --- }We construct a simple model for a purely dissipative, Markovian quantum many-body system consisting of $\ell$ qubits. The Hilbert space dimension is $N=2^\ell$, and the operator space is spanned by all $N^2=4^\ell$ normalized Pauli strings 
\begin{equation}
    S_{\mu} = N^{-1/2} \sigma_{\mu_1} \otimes \sigma_{\mu_2} \otimes\dots \otimes \sigma_{\mu_\ell}, \quad \mu_i\in\{ 0,x,y,z \},
    \label{paulistring}
\end{equation}
where $\sigma_0=\eye$, and $\sigma_{x,y,z}$ are the Pauli matrices. Dissipation is generated by a set of $k$-local jump operators given by $k$-local Pauli strings such that the number of non-identity Pauli matrices in the string is at most $k$. That is, for $k$-local $S_{\mu}$ we have $\sum_{i=1}^{l} (1-\delta_{\mu_i,0}) \leq k$. We will focus on the physically relevant case of two-body dissipative interactions, including one qubit and two qubit ($k=2$) dissipation channels, yielding $N_L=3\ell + 9 {{\ell}\choose{2}}$ jump operators. 

The dynamics of the density matrix $\rho$ is governed by the purely dissipative Liouvillian \cite{Gardiner2004} defined in terms of
a Kossakowski matrix $K_{\mu\nu}$ which encodes the nontrivial couplings between the dissipation channels and is randomly sampled from the ensemble of \emph{positive semidefinite} matrices. The sums $\mu,\nu$ run over the $N_L$ jump operators $L_\mu=S_\mu$, given by $k$-local Pauli strings in Eq. (\ref{paulistring}), 
\begin{equation}
    \mathcal L[\rho]  = \sum_{\mu,\nu=1}^{N_L} K_{\mu\nu} \left[ L_\mu \rho L_\nu^\dagger - \frac 1 2 \{ L_\nu^\dagger L_\mu,\rho\}   \right] .
    \label{eq:lindblad}
\end{equation}
Using the same procedure as in Refs.~\cite{denisov_universal_2019,wang_hierarchy_2020,sommer_many-body_2021}, we generate the i.i.d.\ non-negative eigenvalues of $K$ from a uniform distribution, and normalize them such that $\mathrm{Tr} K = N$. Then, we rotate the basis by a Haar random unitary $U \in \text{CUE}(N_L)$ to yield $K=U^\dagger D U$, where $D$ is the diagonal eigenvalue matrix of $K$. 

In contrast to Ref.~\cite{wang_hierarchy_2020}, we are interested in understanding the effect of a strongly varying dissipation strength across the system. The simplest way to consider this is by splitting the set of jump operators $\{L_\mu\}$ into \emph{strongly} dissipative ones, $\{L_\mu^\text{s}\}$, and \emph{weakly} dissipative ones, $\{L_\mu^\text{w}\}$. This is achieved by defining a set of ``good'' qubits in a system of otherwise ``bad'' qubits: weak jump operators 
are those that contain a non-identity Pauli matrix on a good qubit, so that dissipation happens at a rate scaled by $\sqrt{\alpha}<1$, $L_\mu^\text{w} =\sqrt{\alpha} S_{\mu}$, 
while strong jump operators are still of the form $L_\mu^\text{s} = S_{\mu}$
\footnote{Note that we do not impose a notion of geometry, essentially dealing with a zero dimensional system. A geometry can be added on top as discussed in the supplement of Ref.~\cite{wang_hierarchy_2020}, and adds further complexity, which we do not discuss here to focus on the essential physics of metastability.}.

\textbf{Spectrum of the Liouvillian --- }\begin{figure}[h]
    \centering
    \includegraphics[width=\columnwidth]{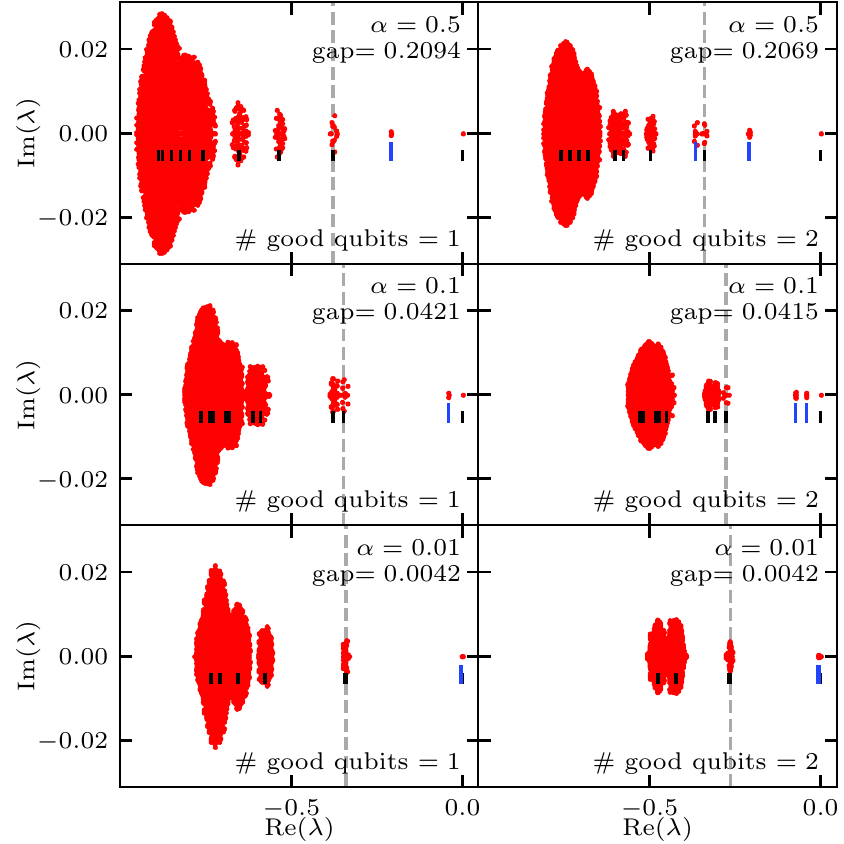}
    \caption{Spectra of random local Liouvillians for $\ell=6$ with one (left column) or two (right column) good qubits as a function of the weak dissipation rate $\alpha$. 
    The spectral gap, given by the magnitude of the real part of the first excited eigenvalue, decreases proportionally with $\alpha$. Bars indicate the eigenvalues of the unperturbed Liouvillian $\mathcal{L}^0$, the starting point of our perturbation theory. Blue bars indicate the position of eigenvalues giving rise to the MM at small $\alpha$.
    For small $\alpha$ there is a separation between metastable eigenmodes and the rest of the spectrum, since all other eigenvalues are have real parts smaller than $\lambda(n_\text{s}=1, n_\text{w}=0)$ given in Eq.~\eqref{eq:pert_eigval}, and indicated by the vertical dashed line in each panel.
    }
    \label{fig:spectra}
\end{figure}
In Fig. \ref{fig:spectra}, we show the complex eigenvalues of a realization of the Liouvillian for $\ell=6$, for one (left) and two (right) good qubits, with two qubit  interactions and one qubit dissipation ($k=2$-local in our definition). The Liouvillian (\ref{eq:lindblad}) is bi-stochastic (as all $L_\mu$ are Hermitian), thus it generically has a single eigenvalue zero with the identity as the unique stationary state, and all other eigenvalues with negative real parts. 

Due to the locality of our model, the spectrum separates into multiple eigenvalue clusters, organized by the locality of their  eigenmodes. If good and bad qubits have the same rate of dissipation ($\alpha=1$), we recover the spectrum of Ref.~\cite{wang_hierarchy_2020}. As we make good qubits better ($\alpha<1$), additional eigenvalue clusters appear. These clusters have a real part proportional to $\alpha$, and are indicated by the blue bars in Fig. \ref{fig:spectra}. For decreasing $\alpha$, these clusters move progressively closer to zero. For small $\alpha$ they combine to form the MM (see below) of long lived states with the slowest relaxation.  
The other clusters also move slightly with $\alpha$, but reach a limiting position well separated from the MM.

Note that in the case of one good qubit, there is a single cluster of three eigenvalues close to zero, while for two good qubits there are two such clusters, one with six and the other with nine eigenvalues. 
To elucidate these spectral properties further, we study these clusters using perturbation theory.

\textbf{Perturbation Theory --- }Due to the physical requirement that the Liouvillian is trace preserving and completely positive, the random $K$ matrix is diagonally dominant \cite{wang_hierarchy_2020}.  It hence has the following properties: the mean of the matrix elements $K_{\mu\nu}$ is $\delta_{\mu,\nu}N/N_{L}$
and the standard deviation  is $N/\left( \sqrt{6}N_{L}^{3/2} \right)$, which can be shown by the central limit theorem and using random matrix theory for $U$.
Hence, we can devise a perturbative treatment by splitting the $K$ matrix as
$K= K_0 + K_1$, where 
$K_0 = \left(N/N_{L}\right)\eye $ is the unperturbed matrix, and $K_1$ a small perturbation which we neglect for now.

We can express the Liouvillian $L$ (or it adjoint) as a matrix in the Pauli string basis \footnote{Note that the normalization is automatic since we defined Pauli strings such that $\text{Tr} S_\mu^2 = 1$.} with matrix elements
$    \mat{L}^{0}_{\mu\nu} = \text{Tr} \left(  S_\mu \mathcal{L} \left[ S_\nu \right]    \right)$.
To leading order off-diagonal matrix elements vanish, and by separating the expressions for weak and strong channels we get for the diagonal elements and thus eigenvalues to leading order,
\begin{equation}
	     \mat{L}^{0}_{\mu\mu} =-\frac{2}{N_L} \left( \alpha N^{\text{w}}_{\mu} + N^{\text{s}}_{\mu} \right),
	      \label{eq:perturbation}
\end{equation}
where $N^{\text{w}}_{\mu}$ and $N^{\text{s}}_{\mu}$ are the numbers of weak and strong jump operators, respectively, that anticommute with $S_{\mu}$. 
The number $n_{\text{w}}$ and $n_{\text{s}} $ of non-identity Pauli matrices on good and bad qubits determine the above.  
For the 2-local case with $\ell_{\text{w}}$ good qubits, we obtain from Eq. \eqref{eq:perturbation} 
\begin{equation}
    \begin{split}
        \lambda(n_s, n_w)
    = &-\frac{2}{N_{L}} \Big[ 6n_{\text{s}}\ell + 6n_{\text{s}}\ell_{\text{w}}(\alpha - 1)  \\
        &  - 4n_{\text{s}}^{2} + \alpha(6n_{\text{w}}\ell - 8 n_{\text{w}} n_{\text{s}} -4 n_{\text{w}}^{2}) \Big].
\end{split}
    \label{eq:pert_eigval}
\end{equation}
Since there are many Pauli strings with the same numbers of  $n_\text{w}$ and $n_\text{s}$, each eigenvalue is highly degenerate. There is a unique steady state, $n_\text{w}= n_\text{s}=0$, corresponding to the identity. 

Including the small perturbation $K_1$ lifts the degeneracy of the eigenvalues, and gives them small imaginary parts. To see this, we diagonalize the Liouvillian with $K=K_1$ inside each degenerate subspace. Lowly degenerate eigenvalues, which are well separated from the rest of the spectrum, develop into the clusters observed in Fig. \ref{fig:spectra}, while for eigenvalues close to others and with high degeneracy the separation does not survive, and the perturbation theory breaks down in these cases (see \cite{SM} for a detailed discussion of this for large systems). For small $\alpha$, the perturbation theory is excellent and yields well separated eigenvalue clusters close to the steady state, as can be seen in Fig. \ref{fig:spectra}. Each eigenvalue cluster in this case is centered around the unperturbed eigenvalue $\lambda(n_\text{s}, n_\text{w})$, indicated by black ($n_\text{s}>0$) and blue ($n_\text{s}=0$) bars, as predicted from our perturbation theory.

Further inspection of Eq. \eqref{eq:pert_eigval} reveals that eigenvalues corresponding to observables with only identities on bad qubits (i.e. $n_\text{s}=0$) are proportional to $-\alpha$, 
while any observable with a non-identity on a bad qubit picks up a constant offset and thus generically has a much faster decay rate. This is what makes up the MM: eigenvalues with real parts proportional to $-\alpha$ are close to zero for small $\alpha$, and well separated from the rest of the spectrum. They are centered around
$\lambda(n_\text{s}=0, n_\text{w}) = -\frac{4\alpha}{N_L} \left( 3 n_\text{w} \ell - 2 n_\text{w}^2 \right),$
which means that for one good qubit we get one eigenvalue cluster (since $n_\text{w}$ can only be either zero or one), and for $\ell_\text{w}$ good qubits, we get $\ell_\text{w}$ separate eigenvalue clusters with eigenvalues proportional to $-\alpha$. The remaining eigenvalues are always smaller than 
$    \lambda(n_\text{s} = 1, n_\text{w}=0) = -\frac{4}{N_L} \left( 3(\ell-\ell_\text{w}) - 2 + 3 \ell_\text{w} \alpha\right),$
indicated by the vertical dashed line in Fig. \ref{fig:spectra}. This sets the separation between the MM and the rest of the spectrum, and thus the relaxation timescale of an arbitrary initial state towards the MM before relaxation to the steady state happens on a timescale $\propto 1/\alpha$.

\textbf{Metastable manifold --- }The existence of eigenvalues with small real parts, which are well separated from the rest of the spectrum for small $\alpha$ gives rise to {\em metastability} \cite{macieszczak_towards_2016,rose2016,rose2020hierarchical,macieszczak_theory_2021}. The evolution $\rho(t) = \mathrm{e}^{t \mathcal L} \rho_0$ of any initial state $\rho_0$ can be written in terms of the eigenvalues $\lambda_m$ and right eigenmatrices $R_m$ of the Liouvillian, 
\begin{equation}
    \rho(t) = R_0 + \sum_{m=1}^{M} \mathrm{e}^{\lambda_m t} c_m R_m +  \sum_{m=M+1}^{4^\ell} \mathrm{e}^{\lambda_m t} c_m R_m,
    \label{eq:evolution}
\end{equation}
where we have split the contribution of the $M$ eigenvalues with largest real parts from the rest of the spectrum, and were the coefficients $c_m$ are given by $c_m=\mathrm{Tr}(L_m \rho_0)$, $L_m$ being the left eigenmatrices. For a large spectral separation, there is a wide range of times for which the modes $m>M$ have already decayed and can be neglected above giving 
\begin{equation}
\rho(t) \approx R_0 + \sum_{m=1}^{M} \mathrm{e}^{\lambda_m t} c_m R_m .
\label{eq:MMdyn}
\end{equation}
This is the metastable regime where dynamics is approximately restricted to the lower dimensional MM. The valid combinations of $c_i$ classify a MM as either classical or quantum \cite{macieszczak_towards_2016,macieszczak_theory_2021}. 
A MM is called \emph{classical} if there exists a basis of density matrices $\tilde{\rho}_i$ so that any state in the MM is a positive linear combination
$    \rho(t)\approx\sum_{i=1}^mp_i\tilde{\rho}_i$ with
 $0\leq p_i \leq 1$. In this case the MM is a simplex, analogous to the manifold of probability distributions, with the $p_i$ the probabilities of being in each metastable phase $\tilde{\rho}_i$, and the long time dynamics can be cast as a classical Markov jump process between these phases. 
When such basis does not exist the MM is said to be {\em quantum}.

At the level of the perturbative calculation above, we can read off the eigenmatrices $R_m$ and $L_m$ ($m \leq M$)  forming the MM. For one good qubit ($\ell_\text{w}=1$), we have three eigenvalues with $n_\text{w}=1, n_\text{s}=0$, and the matrices are the three Pauli strings with a non-identity on the good qubit and the identity. For two good qubits, we obtain two eigenvalue clusters in the MM, which remain well separated even for large $\ell$ (cf.\ discussion in \cite{wang_hierarchy_2020} and \cite{SM}): one is formed by the six one qubit Pauli strings with one identity on one of the good qubits ($n_\text{w}=1$), and the other by the nine two qubit Pauli strings with nonidentities on both good qubits ($n_\text{w}=2$). 
Perturbation theory thus suggests that the MM is quantum, since it is invariant under the action of SU(2) operators on the slow sites. However, this assumption might fail when we take into account the full random $K$ matrix, the additional corrections allowing for a classical manifold to form. We now test this numerically.

To test for classicality of the MM we apply the algorithmic approach of Refs.~\cite{rose2020hierarchical,macieszczak_theory_2021}
which tries to systematically find the best possible simplex from spectral data of the Liouvillian.
Accuracy is measured by a bound on the average distance of metastable states outside this optimal simplex.
This bound follows by noting that given some basis $\tilde{\rho}_m$ $(m=0,\ldots,M)$ of the MM \footnote{The states $\tilde{\rho}_m$ are called `extreme metastable states' \cite{macieszczak_theory_2021}. They define the endpoints of the simplex that best approximates a classical MM, and all metastable states are probabilistic mixtures of them.}.
They correspond to the "metastable phases" that coexist during the long transient regime of metastability, in , there exists a unique dual basis $\tilde{P}_m$ with normalization chosen as $\Tr(\tilde{P}_m\tilde{\rho}_{m'})=\delta_{mm'}$.
Therefore, the coefficients for a state $\rho$ projected to the MM are given by $p_m=\Tr(\tilde{P}_m\rho)$, bounded by the maximum and minimum eigenvalues of $\tilde{P}_m$.
These eigenvalues reside between $0$ and $1$ if the MM is exactly a simplex, and thus classical.
How far any eigenvalues $\lambda_j^{(P_m)}$ of the $\tilde{P}_m$ are outside of this range defines a classicality measure \cite{macieszczak_theory_2021}
\begin{align}\label{eq:classicality}
    \mathcal{C}=\frac{1}{2^{N}}\sum_{m=0}^M\sum_{j=1}^{2^N}\max\left[-\lambda_j^{(P_m)},0\right] .
\end{align}
Since an exactly classical MM has vanishing $\mathcal{C}$, the more $\mathcal{C}$ departs from zero the further away from classical the MM is. 

\begin{figure}
    \includegraphics[width=\linewidth]{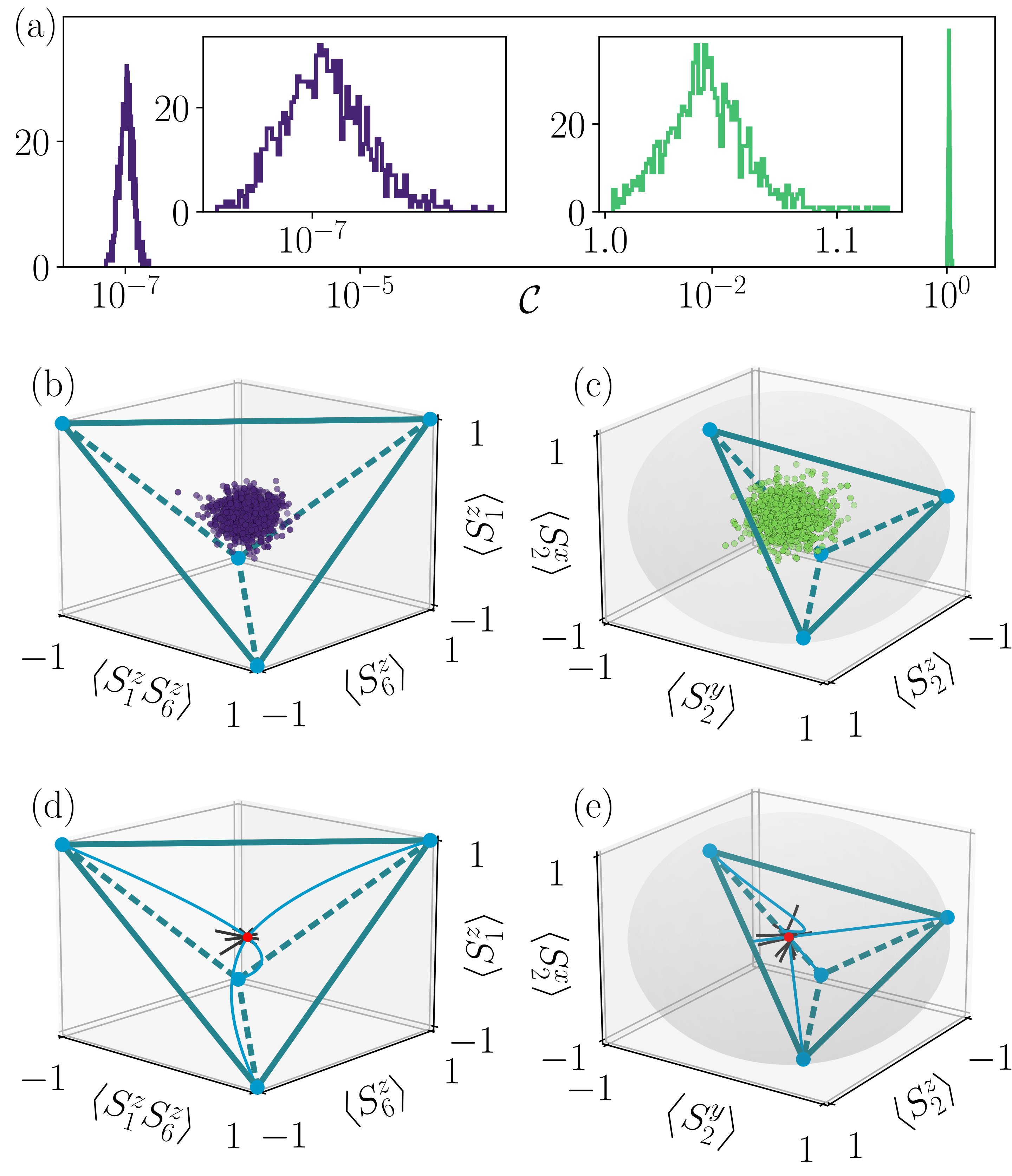}
    \caption{(a) Histograms of the classicality $\mathcal{C}$ for an ensemble of 1000 disorder realizations of $K$, for the quantum MM case (green, right) and the classical MM case (purple, left), for $l=6$. 
(b) Simplex that best approximates the MM in the classical MM case for one disorder realization (blue dots indicate the extreme metastable phases \cite{Note3}, lines the edges of the simplex, dashed lines are behind the volume of the simplex) in which approximately all metastable states are contained if the MM is classical, in the basis of Pauli strings with Z Pauli matrices on the slow sites $1$ and $6$ [$S^z_1 = S_{z00000}, S^z_6 = S_{0000z}, S^z_1 S^z_6 = S_{z0000z}$ in the notation of Eq.~(\ref{paulistring})]. Dots (purple) are projections of a set of random initial states onto the MM, plotted according to the expectation value of the three observables. All states sampled fall within the simplex, as expected for a classical MM. 
(c) Same for the quantum MM case, now in terms of Pauli strings with a non-trivial Pauli matrix on the slow site [$S^x_2 = S_{0x0000}, S^y_2 = S_{0y0000}, S^z_2 = S_{0z0000}$]. Projections of random initial states (green) escape the simplex, as the MM is quantum (the shaded Bloch sphere). 
(d,e) Projections of the time-evolution for long times of the metastable phases (blue curves) and of a set of random initial states (black curves) within the MM towards stationarity (red dot), for the classical MM case (left panel) and quantum MM case (right panel).}
\label{fig:metastability}
\end{figure}

With this procedure we construct the simplex approximation to the MM for a set of 1000 realizations of the disorder matrix $K$, showing a histogram of $\mathcal{C}$ in Fig.~\ref{fig:metastability}\textcolor{blue}{(a)} (green).
We see that the manifold is never classical, $\mathcal{C} \gtrsim 1$, in the disorder realizations we consider.
To illustrate this visually, for one sample realization we plot the projections of random pure states on to the metastable manifold against the expectation values of Pauli operators on the slow site in Fig.~\ref{fig:metastability}\textcolor{blue}{(c)}: we see that many of these projections fall outside the optimal simplex, and indication that the MM is not classical. 
In Fig.~\ref{fig:metastability}\textcolor{blue}{(e)}, we evolve a few metastable states as they converge towards the stationary state, seeing that some spend time outside the simplex (but still within the quantum MM).

The quantum nature is apparently robust in this random matrix model as suggested by perturbation theory.
To obtain a classical MM, we slow only certain Pauli operators on the good qubits. 
For example, we multiply by $\sqrt{\alpha}$ only those jump operators that have $X$ or $Y$ Pauli matrices on the $\ell_w=2$ good qubits, but not those with $Z$. In the perturbation theory, this results in only the $Z$ operators on  good qubits commuting with all rapidly relaxing operators in Eq.~\eqref{eq:perturbation}. In this case, the MM is thus made up of 4 operators: the identity, the $Z$ operator on each good qubit, and the product of $Z$ operators on both good qubits. The algorithms of \cite{rose2020hierarchical,macieszczak_theory_2021} yield an extremely accurate simplex approximation to the MM, confirming that it is effectively classical as shown in Fig.~\ref{fig:metastability}\textcolor{blue}{(a)}.
This is visualized by projecting a set of random states onto the slow-mode eigenspace in Fig.~\ref{fig:metastability}\textcolor{blue}{(b)}, locating them well within the simplex.
Further, as shown in Fig.~\ref{fig:metastability}\textcolor{blue}{(d)}, the long-time evolution of a set of these metastable states (black), or the metastable phases (blue) remain within the simplex at all times.

\textbf{Metastable dynamics --- }\begin{figure}[h]
    \centering
    \includegraphics{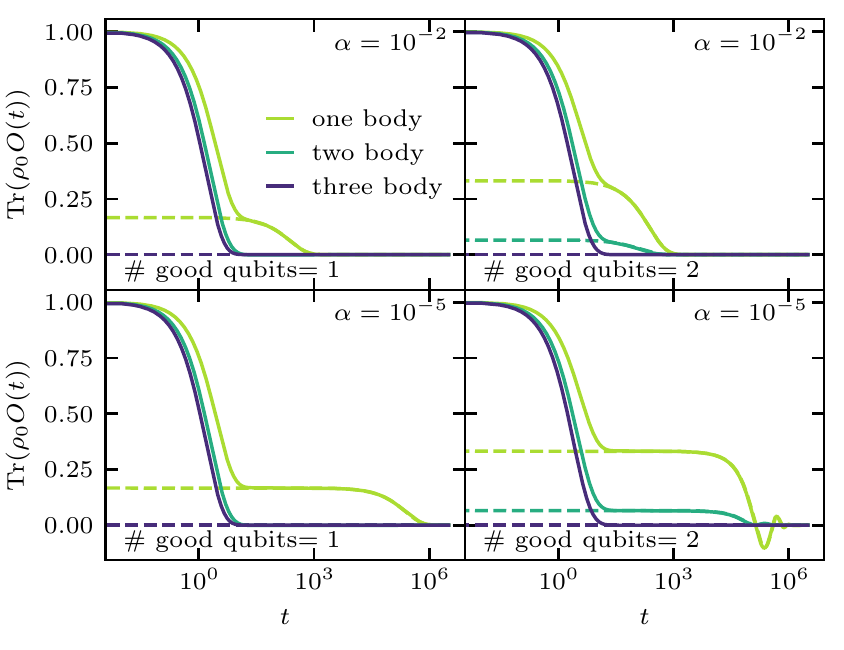}
    \caption{Comparison of the exact dynamics (solid lines) with the dynamics projected onto the MM (dashed lines) for $\ell = 7$, for qubits with one (left column) and two good (right column) qubits at two different dissipation rates $\alpha=10^{-2}, 10^{-5}$. We show expectation values of three observables with different localities, each prepared by a random linear superposition of all the Pauli matrices of the corresponding locality. Observables supported by the bad qubits vanish rapidly, and the long-time dynamics on the good qubits coincide with the effective dynamics. Note that for a MM to display, the locality of the observable has to be smaller or equal to the number of the good qubits. }
    \label{fig:dyn}
\end{figure}
Using Eq. (\ref{eq:evolution}), we can calculate the evolution of observables at any time. To consider generic  initial states we choose $\rho_{0}$ as a random linear superposition of the full Hilbert space. Figure \ref{fig:dyn} (full lines) shows the time evolution of  observables with different locality properties (non-trivial Pauli strings of different lengths $k$). Note the appearance of plateaus in the relaxation curves, specifically for the shorter Pauli strings which have a larger overlap with the matrices $R_m$ that define the MM. 

After a fast transient, dynamics is confined to the MM. 
The approximate dynamics is then obtained by projecting both the initial state and the observables onto it, and solving Eq.~(\ref{eq:MMdyn}). The dashed curves in 
Fig. \ref{fig:dyn} show the corresponding results: the effective dynamics captures the long-time behavior accurately, showing that metastability implies dimensional reduction from the whole Hilbert space to the MM.

\textbf{Conclusion --- }Starting from a random local and purely dissipative Liouvillian, we have defined a random matrix model for generic metastability in open quantum systems relevant for strongly varying dissipation timescales in quantum computers. We find that a separation of dissipation timescales induces the presence of a metastable manifold to which initial states relax, before the evolution to the steady state occurs at much longer times. If the dissipation on good qubits does not further single out certain Pauli operators, we show that the metastable manifold is generically quantum, while further structure can lead to classical manifolds instead.

\begin{acknowledgments}
This work was supported in part by the Deutsche Forschungsgemeinschaft
through SFB 1143 (Project-id 247310070) and the cluster of excellence 
ML4Q (EXC 2004). JPG and DCR acknowledge financial support from EPSRC Grant no. EP/R04421X/1 and the Leverhulme Trust Grant No. RPG-2018-181. 
\end{acknowledgments}

\bibliography{references}

\newpage

\appendix

\section{SUPPLEMENTAL MATERIAL: Scaling of eigenvalue cluster width and distance}

The perturbation theory employed in the main text predicts the positions $\lambda(n_\text{s}, n_\text{w})$ of the eigenvalue clusters for a size $\ell$ of the system and any number of good qubits $\ell_\text{w}$. The perturbation theory is valid if the eigenvalue clusters are well separated, in which case the degeneracy of the eigenvalues $\lambda(n_\text{s}, n_\text{w})$ is lifted such that the resulting eigenvalue clusters do not overlap.
Since the number of different eigenvalues grows with system size $\ell$ (we consider the case of fixed $\ell_\text{w}$ here), and the number of eigenvalues contained in each cluster grows as well, we expect that for large sizes the perturbation theory only works for the best separated clusters, in particular the MM.

\begin{figure}[h]
    \centering
    \includegraphics{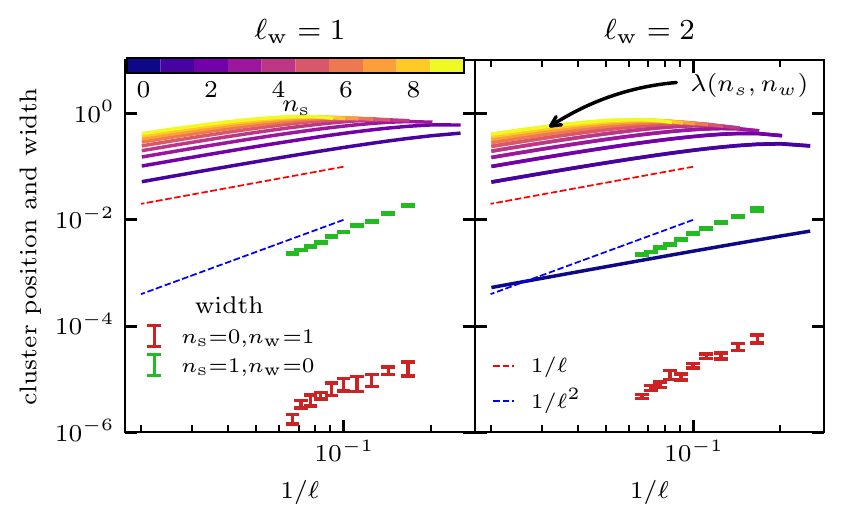}
    \caption{Scaling of the position (solid color lines) and width (errorbars) of eigenvalue clusters for one ($\ell_\text{w}=1$, left) and two ($\ell_\text{w}=2$, right) good qubits and $\alpha=0.01$. The width of the two eigenvalue clusters with largest real parts was obtained by degenerate perturbation theory, diagonalizing the adjoint Liouvillian in the corresponding degenerate subspace of the unperturbed problem. Errorbars stem from disorder averaging over several dozen realizations of $K$.  }
    \label{fig:scaling}
\end{figure}

To test this, we show results from our degenerate perturbation theory in Fig. \ref{fig:scaling}, for one ($\ell_\text{w}=1$, left panel) and two ($\ell_\text{w}=2$, right panel) good qubits as a function of total system size $\ell$. Solid lines show the eigenvalues $\lambda(n_\text{s}, n_\text{w})$ of the unperturbed problem, i.e. the centers of the expected eigenvalue clusters. These generically scale as $1/\ell$ (red dashed lines). The color of the lines encodes the number of non-identity operators on bad qubits $n_\text{s}$, which is the dominant ingredient for the cluster position, while the number of nonidentities on good qubits $n_\text{w}$ only contributes a finestructure for small $\alpha$.

The positions of the predicted eigenvalue clusters are compared to the width of the eigenvalue clusters [$|\text{max} \text{Re}(\lambda_i) - \text{min}\text{Re}(\lambda_i)|$, with eigenvalues $\lambda_i$] after applying degenerate perturbation theory to lift the degeneracy. 
The width of the first (i.e. with largest real parts of their eigenvalues) two clusters is shown by red ($n_\text{s}=0, n_\text{w}=1$) and green ($n_\text{s}=1, n_\text{w}=0$)  errorbars in Fig. \ref{fig:scaling}, revealing a scaling of the width of the eigenvalue clusters $\propto 1/\ell^2$.

Therefore, we conclude that the eigenvalue clusters forming the MM remain well separated from the rest of the spectrum even in the limit of $\ell\to \infty$, if the number of good qubits $\ell_\text{w}$ is fixed. This separation persists, because the width of the eigenvalue clusters decreases faster with system size compared to the distance to neighboring clusters.

\end{document}